\documentclass{article}

\usepackage{amsmath,amssymb,graphicx}

\pdfoutput=1

\begin{document}

\begin{center}

{\Large Exact solution of a cluster model with \\ next-nearest-neighbor interaction}

\vspace{0.6cm}



Yuji Yanagihara$^{*}$, Kazuhiko Minami$^{\dagger}$



\vspace{0.6cm}

$^{*}$General Education, Fukui College, National \\ Institute of  Technology, Fukui 916-8507, Japan \\

$^{\dagger} $Graduate School of Mathematics, Nagoya University, \\ 
Furo-cho, Chikusa-ku, Nagoya, Aichi, 464-8602, Japan 

\end{center}

\begin{abstract} 
A one-dimensional cluster model with next-nearest-neighbor interactions
and two additional composite interactions is solved; 
the free energy is obtained and a correlation function is derived exactly. 
The model is diagonalized by a transformation obtained automatically from its
 interactions,  which is an algebraic generalization of the Jordan-Wigner transformation. 
The gapless condition is expressed as a condition on the roots of a cubic equation, 
and the phase diagram is obtained exactly. 
We find that the distribution of roots for this 
algebraic equation determines the existence of long-range order, 
and we again obtain the ground-state phase diagram. 
We also derive the central charges of the corresponding CFT.   
Finally, we note that our results are universally valid for an infinite number of 
solvable spin chains whose interactions obey the same algebraic relations. 
\end{abstract}


%
%
%
%

\section{Introduction} 

Quantum spin models have been widely investigated as a basic theme 
in statistical physics, mathematical physics, and condensed matter physics. 
In one-dimension, there exist many solvable systems, 
and there also exist general relationships with 
two-dimensional classical systems. 
Specifically, a one-dimensional XY-model was introduced in [1]-[4] 
and was later found to be equivalent to the two-dimensional rectangular Ising model[5].
In the 2000s, the cluster models attracted wide attention.
The ground state of this model is called the cluster state, 
which is a candidate of a resource for measurement-based quantum computation(MBQC)[6][7][8]. 
The one-dimensional cluster model was first introduced and solved by Suzuki[5] 
and has been investigated by numerous researchers[9]-[22].
A new method of fermionization has recently been introduced[23], and an infinite number of 
new solvable models have been reported in [23]-[25]. 
In this method, the transformation that diagonalizes the system is obtained automatically 
from the interactions of the model.
The Hamiltonian is diagonalized using only algebraic relations of the interactions, and therefore, 
the equivalences of the models are immediately understood.
In the case of the XY model, the transformation results in the
Jordan-Wigner transformation[26][2]; thus, this new transformation can be 
regarded as the algebraic generalization of the Jordan-Wigner transformation. 
This algebraic method has been developed 
into a graph theoretical method of fermionization[27],  
in which transformations of operators are expressed by deformations of graphs, 
and the kernels of their adjacency matrices provide conserved quantities 
of the systems. 
In this paper, we investigate the cluster model 
with next-nearest-neighbor interactions
and two additional composite interactions. 
The Hamiltonian is
 \begin{eqnarray}
    - \beta \mathcal{H} 
   &=& 
      K_0 \sum_{j = 1}^N  \sigma_j^x \sigma_{j+1}^z \sigma_{j+2}^x
     +K_1 \sum_{j = 1}^N  \sigma_j^x 1_{j+1} \sigma_{j+2}^x 
   \nonumber \\
   & &
     +K_2 \sum_{j=1}^N  \sigma_j^x \sigma_{j+1}^x  \sigma_{j+2}^z \sigma_{j+3}^x\sigma_{j+4}^x
     +K_{-1} \sum_{j=1}^N  \sigma_j^x \sigma_{j+1}^y 1_{j+2} \sigma_{j+3}^y \sigma_{j+4}^x
   , 
   \label{Hamiltonian} 
 \end{eqnarray}
where $ 1_j $ is the identity operator at site $j$.
%
%
%
When $ K_2 = K_{-1} = 0 $, this model becomes the cluster model with
 next-nearest-neighbor interactions 
and found to be basically equivalent [24] to the XY chain.
However, the model (\ref{Hamiltonian}) cannot be diagonalized 
by the Jordan-Wigner transformation.
Our formula can diagonalize this Hamiltonian.
The formula in this paper is summarized as follows.
Let the number of sites be  $N = 2M$ , where $M$ is even, and let us assume a cyclic  boundary condition
    $ \sigma_{N+i}^k  =  \sigma_i^k $ 
    ( $ k = x , y , z $ ).
Let us consider two series of operators 
    $ \{ \eta_j \} $ and $ \{ \zeta_j \} $,
which are defined as 
  \begin{eqnarray}  
      \eta_{2j-1} = \sigma_{2j-1}^x \sigma_{2j}^z  \sigma_{2j+1}^x  , \ \ 
      \eta_{2j} = \sigma_{2j}^x  1_{2j+1}   \sigma_{2j+2}^x  , \ \ ( 1 \leq j \leq \frac{N}{2} - 1)
    \notag \\  
      \eta_{N-1} = \sigma_{N-1}^x \sigma_{N}^z \sigma_{1}^x ,\ \  
      \eta_{N} = \sigma_{N}^x 1_{1} \sigma_{2}^x
     ,
   \label{op1} 
  \end{eqnarray}
and
  \begin{eqnarray}
      \zeta_{2j-1} = \sigma_{2j}^x  \sigma_{2j+1}^z  \sigma_{2j+2}^x , \ \
      \zeta_{2j} = \sigma_{2j+1}^x   1_{2j+2}   \sigma_{2j+3}^x , \ \ ( 1 \leq j \leq \frac{N}{2}-1 ) 
    \notag \\ 
      \zeta_{N-1} = \sigma_{N}^x  \sigma_1^z  \sigma_{2}^x , \ \ 
      \zeta_{N} = \sigma_{1}^x 1_2 \sigma_{3}^x
      .
   \label{op2} 
  \end{eqnarray}
Note that the operators $ \eta_j $ and $ \zeta_k $ are the interactions found 
in the Hamiltonian (\ref{Hamiltonian}), and the operators $\eta_j $ and $\zeta_k $ 
commute with each other for all $j$ and $k$.
Next, following equation (2.9) in [23], we introduce transformations 
 $ \varphi_1(j) $, $ \varphi_2(j) $, and $ \varphi_3(j) $, $ \varphi_4(j) \ $( $ 1 \leqq j \leqq M $ )
as
  \begin{gather}
     \varphi_1(j) = \frac{1}{\sqrt{2}} e^{i \frac{ \pi }{2} (k-1)} \eta_0 \eta_1 \cdots \eta_k
        \ \
         ( k=2j-2), \notag \\
     \varphi_2(j) = \frac{1}{\sqrt{2}} e^{i \frac{ \pi }{2} (k-1)} \eta_0 \eta_1 \cdots \eta_k 
        \ \ 
        ( k=2j-1), \notag \\
     \varphi_3(j) = \frac{1}{\sqrt{2}} e^{i \frac{ \pi }{2} (k-1)} \zeta_0 \zeta_1 \cdots \zeta_k
        \ \ 
        ( k=2j-2), \notag  \\
     \varphi_4(j) = \frac{1}{\sqrt{2}} e^{i \frac{ \pi }{2} (k-1)} \zeta_0 \zeta_1 \cdots \zeta_k
        \ \ ( k=2j-1)
      ,
   \label{a} 
  \end{gather}
where
  \begin{equation}
       \eta_0 = i  \sigma_1^x  \sigma_2^x 
     \ \ \ {\rm and } \ \ \ 
       \zeta_0 = i  \sigma_2^x \sigma_3^x
  \label{b} 
  \end{equation}
are the initial operators. 
Then, the operators $\varphi_{l}(j)$ satisfy 
  \begin{eqnarray}
     \{\varphi_{l}(j), \varphi_{m}(k)\} &=& \delta_{lm} \delta_{jk} \hspace{1.4cm} (l, m=1, 2), \nonumber \\
     \{\varphi_{l}(j), \varphi_{m}(k)\} &=& \delta_{lm} \delta_{jk} \hspace{1.4cm} (l, m=3, 4), \nonumber \\ 
     \varphi_{l}(j) \varphi_{m}(k) &=& \varphi_{m}(k) \varphi_{l}(j) \hspace{0.6cm} (l=1, 2, \: m=3, 4), 
  \end{eqnarray}
for all $j$ and $k$. 
Therefore, the operators $ \varphi_l(j) $ form two series of Majorana fermions.
The Hamiltonian (\ref{Hamiltonian}) is expressed as the sum of two-body products of $ \varphi_l(j) $.
Hence, the Hamiltonian (\ref{Hamiltonian}) can be diagonalized. 
The transformations from $ \eta_j $ to $ \varphi_1 $ and $ \varphi_2 $
 ( $\zeta_j$ to $ \varphi_3 $ and $ \varphi_4 $ ) are clearly different from
the Jordan-Wigner transformation as shown in (12)-(15). 
From the transformations (\ref{a}), 
we can derive the free energy and a correlation function exactly.

Next, let us consider the algebraic equation
 \begin{equation}
   \alpha_2 z^3 + \alpha_1 z^2 - z + \alpha_{-1} =0
   ,  
  \label{equ1} 
 \end{equation}
where $ \alpha_{-1} = K_{-1}/K_0 $, $ \alpha_1 = K_1/K_0 $, $\alpha_2 = K_2/K_0 $, 
and also consider three subsets of the complex plane
 \begin{equation}
    C_u  =  \{ z \in \mathbb{C} \ | \ |z| = 1 \}, \ \ 
    D_O  =  \{ z \in \mathbb{C} \ | \ |z| > 1 \}, \ \  
    D_I  =  \{ z \in \mathbb{C} \ | \ |z| < 1 \}.
 \end{equation}
We find that there is no energy gap above the ground state 
if and only if at least one root of equation (\ref{equ1}) belongs to the unit circle $C_u$.
We also find that the existences of long-range orders are classified by locations of the roots of the cubic equation (\ref{equ1}). 
The boundary of the phase 
in terms of the long-range order satisfies the gapless condition;
therefore, we obtain a consistent phase diagram 
from two different procedures.
We also determine the universality classes of these phase transitions by deriving the central charges of the corresponding CFT. 
We also find that the phase diagram has a symmetry under a shift 
of indices from $ j $ to $ j+1 $ for $ K_j $.
These results are obtained using only the algebraic relations of the interactions. 
Hence, the results we obtained are universally valid for models 
consisting of the interactions 
that obey the same algebraic relations.  

In section 2, we diagonalize the Hamiltonian (\ref{Hamiltonian}) exactly 
by applying the transformation (\ref{a}) and obtain the free energy.
In section 3, we consider the gapless condition and obtain a corresponding phase diagram.
In section 4, first a correlation function is introduced, 
and the asymptotic limit is obtained exactly.
Next, we classify the existences of long-range orders in terms of locations
of the roots of equation (\ref{equ1}), and again, we obtain the same phase diagram.
In section 5, we derive the central charges from the finite size behavior of the 
energy spectrum. 
In section 6, we consider the symmetry of the phase diagram, and 
illustrate that the results obtained in this paper 
are valid for an infinite number of Hamiltonians 
that satisfy our condition. 
\section{Diagonalization and the free energy}      

In this section, we diagonalize the Hamiltonian (\ref{Hamiltonian})
and derive the free energy.
The operators  
 $ \eta_j $ and $ \eta_k $ (  $\zeta_j $ and  $\zeta_k $ )
 are  called
 $ adjacent  $  if  $ (j,k)=(j,j+1) $
 $( 1 \leq j \leq N -1 ) $ or $ (j,k) = ( N , 1 ) $ .
Then, the operators 
   $ \{ \eta_j \} $ 
in (\ref{op1}) satisfy the relations
 \begin{equation}
     \eta_j \eta_k =
       \begin{cases}
             - \eta_k \eta_j   & \eta_j \ {\rm and } \ \eta_k \ {\rm are } \ adjacent \\
                \eta_k \eta_j   & \eta_j \ {\rm and } \ \eta_k \ {\rm are \ not } \ adjacent \\
                                  1 & j=k,
      \end{cases}
    \label{2}  
 \end{equation}
and $ \{ \zeta_j \} $ satisfy the same relations replacing $\eta_j $ by $ \zeta_j $ in $ (\ref{2}) $.
Note that
 \begin{equation}
         (\eta_1 \cdots \eta_k )^2 = 
         (\zeta_1 \cdots \zeta_k )^2 = (-1)^{k-1} \ \ ( k < N).
  \label{alt} 
 \end{equation}
Other interactions in (\ref{Hamiltonian}) are obtained from $ \eta_j $ and $ \zeta_k $ as 
 \begin{gather}
       \eta_{2j-1} \eta_{2j} \eta_{2j+1} 
           =
       \sigma_{2j-1}^x \sigma_{2j}^y 1_{2j+1} \sigma_{2j+2}^y \sigma_{2j+3}^x, 
    \notag \\
       \eta_{2j} \eta_{2j+1} \eta_{2j+2} 
          =
       -\sigma_{2j}^x \sigma_{2j+1}^x \sigma_{2j+2}^z \sigma_{2j+3}^x \sigma_{2j+4}^x, 
    \notag \\
       \zeta_{2j-1} \zeta_{2j} \zeta_{2j+1}
          =
       \sigma_{2j}^x \sigma_{2j+1}^y 1_{2j+2} \sigma_{2j+3}^y \sigma_{2j+4}^x,
    \notag \\
       \zeta_{2j} \zeta_{2j+1} \zeta_{2j+2} 
          =
       -\sigma_{2j+1}^x \sigma_{2j+2}^x \sigma_{2j+3}^z \sigma_{2j+4}^x \sigma_{2j+5}^x
   .
  \end{gather}
Then the Hamiltonian (1) is written in terms of $ \eta_j $ and $ \zeta_k $ as
 \begin{eqnarray}
     -\beta \mathcal{H}
               &=&  K_0  \sum_{j=1}^M  ( \eta_{2j-1} + \zeta_{2j-1} ) 
                    + K_1  \sum_{j=1}^M  ( \eta_{2j} + \zeta_{2j} )   
          \nonumber \\
               &-&  K_2  \sum_{j=1}^M  ( \eta_{2j} \eta_{2j+1} \eta_{2j+2} + \zeta_{2j} \zeta_{2j+1} \zeta_{2j+2}  )   
          \nonumber \\
               &+&  K_{-1}  \sum_{j=1}^M (\eta_{2j-1} \eta_{2j} \eta_{2j+1} + \zeta_{2j-1} \zeta_{2j} \zeta_{2j+1}  )
     .
   \label{mo} 
  \end{eqnarray}
Next, let us consider the transformations 
    $ \varphi_k(j) $  ( $ k = 1, 2, 3, 4 $ , $ 1 \le j \le M $ )
introduced in (\ref{a}).
They are written in terms of the Pauli operators as
 \begin{eqnarray}
       \varphi_1(j)
     &=&
        \frac{ 1 }{ \sqrt{2} }  \left( \prod_{\nu=1}^{j-1} 1_{2\nu-1} \sigma_{2\nu}^z \right) \sigma_{2j-1}^x \sigma_{2j}^x,         
   \label{star1} \\  
       \varphi_2(j)
     &=& 
        \frac{ 1 }{ \sqrt{2} }  \left( \prod_{\nu=1}^{j-1} 1_{2\nu-1}\sigma_{2\nu}^z \right) 1_{2j-1} \sigma_{2j}^y \sigma_{2j+1}^x,
   \label{star2} \\ 
       \varphi_3(j) 
    &=&
        \frac{ 1 }{ \sqrt{2} }  \left( \prod_{\nu=1}^{j-1} 1_{2\nu} \sigma_{2\nu+1}^z \right) \sigma_{2j}^x \sigma_{2j+1}^x,         
   \label{star3} \\  
       \varphi_4(j)
     &=& 
        \frac{ 1 }{ \sqrt{2} }  \left( \prod_{\nu=1}^{j-1} 1_{2\nu}\sigma_{2\nu+1}^z \right) 1_{2j} \sigma_{2j+1}^y \sigma_{2j+2}^x
            \ \ ( j = 1, 2, 3 \cdots ).
   \label{star4}  
 \end{eqnarray}
They are clearly different from the Jordan-Wigner transformation. 
The initial operators 
      $ \eta_0 = i  \sigma_1^x  \sigma_2^x  $ 
and
      $ \zeta_0 = i  \sigma_2^x \sigma_3^x $
are not necessary to diagonalize the Hamiltonian, but introduced to avoid 
boundary terms in (\ref{star1}) - (\ref{star4}).
The initial operators satisfy the relations 
 \begin{gather}
     \eta_0^2  =  -1 , \ \zeta_0^2 = -1 ,  \  \eta_0 \zeta_0 = \zeta_0 \eta_0 ,  
  \notag \\
     \eta_0 \eta_1  =  -\eta_1 \eta_0 ,  \ \  \eta_0 \eta_j = \eta_j \eta_0  \ \  ( 2 \leqq j \leqq N ),  
  \notag \\
     \zeta_0 \zeta_1  =  -\zeta_1 \zeta_0 , \ \ \zeta_0 \zeta_j = \zeta_j \zeta_0  \ \  ( 2 \leqq j \leqq N ). 
 \end{gather}
From (\ref{alt}) and (17), 
 we find that
 \begin{equation}
    \{ \varphi_l(j) , \varphi_m(k) \} = \delta_{lm} \delta_{jk} \ \ 
    ( l,m = 1,2  \ {\rm or } \  l,m = 3,4,  \ \  1 \leqq  j , k \leqq M )
    ,
  \label{anti} 
 \end{equation}
and 
 $ \varphi_l (j) $ ($ l = 1, 2 $) 
and
 $ \varphi_m (k) $ ($ m = 3, 4 $) 
commute with each other. 
Moreover, we find
\begin{eqnarray}
  (+2i)\varphi_{2}(j)\varphi_{1}(j+2)
   &=&\eta_{2j}\eta_{2j+1}\eta_{2j+2}
 \hspace{0.3cm}
     =-\sigma^{x}_{2j}\sigma^{x}_{2j+1}\sigma^{z}_{2j+2}\sigma^{x}_{2j+3}\sigma^{x}_{2j+4},
 \nonumber\\
  (-2i)\varphi_{2}(j)\varphi_{1}(j+1)
   &=&\eta_{2j}
 \hspace{2.1cm}
     =\sigma^{x}_{2j} 1_{2j+1} \sigma^{x}_{2j+2},
 \label{et} \\ 
  (+2i)\varphi_{2}(j)\varphi_{1}(j)
 \hspace{0.7cm}
   &=&\eta_{2j-1}
 \hspace{1.75cm}
     =\sigma^{x}_{2j-1}\sigma^{z}_{2j}\sigma^{x}_{2j+1},
 \nonumber\\
  (-2i)\varphi_{2}(j)\varphi_{1}(j-1)
   &=&\eta_{2j-3}\eta_{2j-2}\eta_{2j-1}
     =\sigma^{x}_{2j-3}\sigma^{y}_{2j-2}1_{2j-1}\sigma^{y}_{2j}\sigma^{x}_{2j+1},
 \nonumber
\end{eqnarray}
and 
\begin{eqnarray}
    (+2i)\varphi_{4}(j)\varphi_{3}(j+2)
  &=&\zeta_{2j}\zeta_{2j+1}\zeta_{2j+2}
 \hspace{0.3cm}
    =-\sigma^{x}_{2j+1}\sigma^{x}_{2j+2}\sigma^{z}_{2j+3}\sigma^{x}_{2j+4}\sigma^{x}_{2j+5},
 \nonumber\\
   (-2i)\varphi_{4}(j)\varphi_{3}(j+1)
  &=&\zeta_{2j}
 \hspace{2.1cm}
    =\sigma^{x}_{2j+1}1_{2j+2}\sigma^{x}_{2j+3},
         \\  
   (+2i)\varphi_{4}(j)\varphi_{3}(j)
 \hspace{0.7cm}
  &=&\zeta_{2j-1}
 \hspace{1.75cm}
    =\sigma^{x}_{2j}\sigma^{z}_{2j+1}\sigma^{x}_{2j+2},
 \nonumber\\
   (-2i)\varphi_{4}(j)\varphi_{3}(j-1)
  &=&\zeta_{2j-3}\zeta_{2j-2}\zeta_{2j-1}
    =\sigma^{x}_{2j-2}\sigma^{y}_{2j-1}1_{2j}\sigma^{y}_{2j+1}\sigma^{x}_{2j+2}.
 \nonumber
\end{eqnarray}
Then the Hamiltonian is written as
 \begin{align}
    -\beta \mathcal{H}  =
 \ \ \ &(-2i) K_{-1}   \sum_{j=1}^M  ( \varphi_2(j) \varphi_1(j-1) + \varphi_4(j) \varphi_3(j-1) )
    \notag \\
             + &(+2i) K_0    \sum_{j=1}^M  ( \varphi_2(j) \varphi_1(j)    + \varphi_4(j) \varphi_3(j) )         
    \notag \\ 
             + &(-2i) K_1    \sum_{j=1}^M  ( \varphi_2(j) \varphi_1(j+1) + \varphi_4(j) \varphi_3(j+1) )
     \notag \\
             + &(-2i) K_2     \sum_{j=1}^M  ( \varphi_2(j) \varphi_1(j+2) + \varphi_4(j) \varphi_3(j+2) )
   ,
  \label{majo} 
 \end{align}
which is the sum of two-body products of the Majorana fermion operators $ \varphi_l(j) $.  
  
For the purpose to check the boundary condition, let us consider the boundary terms
 \begin{eqnarray}
    \varphi_2 (M)  \varphi_1 (M+l) 
        &=& 
      \frac{i}{ \sqrt{2} } ( - \sigma_2^z \sigma_4^z \cdots \sigma_N^z )  
                                  \sigma_N^x \sigma_1^x  \cdot  \varphi_1 (M+l) , 
  \\
    \varphi_4 (M) \varphi_3 (M+l)
        &=&
      \frac{i}{ \sqrt{2} } ( - \sigma_1^z \sigma_3^z \cdots \sigma_{N-1}^z )
                                  \sigma_1^x \sigma_2^x  \cdot  \varphi_3(M+l) .
 \end{eqnarray} 
Then the cyclic boundary condition $ \sigma_{N+i}^k = \sigma_i^k $ yields, for $ l = 1$ and $ 2 $, that
  \begin{equation}
     \varphi_1 (M+l) = 
       \begin{cases}
             \varphi_1 (l) &  \ \sigma_2^z \sigma_4^z \cdots \sigma_N^z = -1 \\
           -\varphi_1 (l) &  \ \sigma_2^z \sigma_4^z \cdots \sigma_N^z = 1
       \end{cases}
   \label{d} 
 \end{equation}
and
  \begin{equation}
     \varphi_3(M+l) = 
       \begin{cases}
          \varphi_3(l) &  \ \sigma_1^z \sigma_3^z \cdots \sigma_{N-1}^z = -1 \\
        -\varphi_3(l) &  \ \sigma_1^z \sigma_3^z \cdots \sigma_{N-1}^z = 1.
       \end{cases}
   \label{e} 
 \end{equation}
    
Next let us introduce the Fourier transformation
 \begin{equation}
     \varphi_l (j)  =  \frac{1}{ \sqrt{M} }  \sum_{ 0 \leq q < \pi } ( e^{iqj} C_l(q)  +  e^{-iqj} C_l^{\dagger}(q) ),
 \end{equation}
where
  \begin{align}
     \{ C_l^{\dagger}(p) , C_m(q)  \}  &=  \delta_{lm} \delta_{pq} ,   
   \notag \\
     \{ C_l^{\dagger}(p) , C_m^{\dagger}(q) \}  &=  \{ C_l(p), C_m(q) \}  =  0 
     \ \  ( l, m = 1, 2 \ {\rm or } \ l, m = 3, 4 ),
  \end{align}
and
 $ C_l^{\dagger} (p) $, $ C_l (p) $  $ ( l = 1 , 2 ) $ 
and
 $ C_m^{\dagger} (q) $, $ C_m (q) $  $ ( m = 3 , 4 ) $
commute with each other.
From (\ref{d}) and (\ref{e}), we find
 \begin{equation}
     q =  \frac{2k}{M} \pi  \ \  ( k = 0 ,1 , \cdots , \frac{M}{2} - 1 )
 \end{equation}
when
 $ \sigma_2^z  \sigma_4^z  \cdots  \sigma_N^z  = -1 $
and when
 $ \sigma_1^z  \sigma_3^z  \cdots  \sigma_{N-1}^z  = -1 $.
We also find
 \begin{equation}
     q =  \frac{2k-1}{M} \pi  \ \  ( k = 1 , 2 , \cdots , \frac{M}{2} )
   \label{bc} 
 \end{equation}
when
 $ \sigma_2^z  \sigma_4^z  \cdots  \sigma_N^z  = 1 $
and when
 $ \sigma_1^z  \sigma_3^z  \cdots  \sigma_{N-1}^z  = 1 $.
Without loss of generality, we can assume (\ref{bc}). 
The Hamiltonian is then expressed as
 \begin{equation}
      - \sum_{ 0 < q < \pi } ( W_{12}(q) + W_{34}(q) )
    ,
   \label{dia1} 
 \end{equation}
where
  \begin{gather}
     W_{12} (q)  =  2i L(q)  C_2 ^{\dagger} (q)  C_1 (q)  +  2i L (q)^{\ast}  C_2(q) C_1^{\dagger} (q) ,  \notag \\
     W_{34} (q)  =   2i L(q)  C_4 ^{\dagger} (q)  C_3 (q)  +  2i L (q)^{\ast}  C_4 (q) C_3^{\dagger} (q) ,
  \end{gather}
and
 \begin{equation}
     L(q) =   K_{-1} e^{-iq}  -  K_0  +  K_1 e^{iq}  +  K_2 e^{2iq} 
     .
  \label{dia2} 
 \end{equation}
Here $ L(q)^{\ast} $ is the complex conjugate of $ L(q) $ .
For each $q>0$, the Hamiltonian is therefore the sum of two commutative operators.
We obtain
 \begin{align}
     A(q)  =   L(q) L(q)^{\ast}  
           &=  K_{-1}^2 + K_0^2 + K_1^2 + K_2^2   -  2( K_{-1} K_0 + K_0 K_1 - K_1K_2 ) \cos q \notag \\
           & \ \ \ \ +2( K_{-1} K_1  -  K_0 K_2 ) \cos2q  +  2K_{-1} K_2 \cos3q ,
 \end{align}
which is real and non-negative.

With respect to the basis
 $ \left| 0 \right \rangle $ ,
 $ C_1^{\dagger}(q) \left| 0 \right \rangle $ ,
 $ C_2^{\dagger}(q) \left| 0 \right \rangle $ ,
 $ C_2^{\dagger}(q) C_1^{\dagger}(q) \left| 0 \right \rangle $
 ,
 $ W_{12}(q) $ is expressed as
  \begin{equation}
    \begin{pmatrix}
       0 & 0 & 0 & 0 \\
       0 & 0 & 2iL(q)^{\ast} & 0 \\
       0 & -2iL(q) & 0 & 0 \\
       0 & 0 & 0 & 0
    \end{pmatrix},
  \end{equation}
and its eigenvalues are found to be  $ 0 , 0 , \pm 2 \sqrt{A(q)} $. 
In the same way, the eigenvalues of $ W_{34}(q) $ are found to be $ 0 , 0 , \pm 2 \sqrt{A(q)} $.

The partition function is obtained as
  \begin{align}
     Z &=  \prod_{0<q<\pi} ( e^0 + e^0 + e^{\Lambda(q)} + e^ {-\Lambda(q) })^2 
   \notag \\
        &=  \prod_{0<q<\pi} ( e^{\frac{1}{2} \Lambda(q) } + e^{ - \frac{1}{2} \Lambda(q) } )^4 ,
  \end{align}
where $ \Lambda(q) = 2 \sqrt{A(q)} $ .
The free energy is
  \begin{align}
     - \beta f  &=  \lim_{ N \to \infty } \frac{ \log Z }{N}  \notag \\
                   &=  \frac{1}{ \pi } 
                           \int_0^{\pi} 
                             \log( e^{ \frac{1}{2} \Lambda(q) } + e^{ - \frac{1}{2} \Lambda(q) } ) d q ,
   \label{f}  
  \end{align}
where we have used that
 $ \displaystyle \Delta q  = \frac{2}{ N/2 } \pi $.

When $ K_2 = K_{-1} = 0 $, the free energy (\ref{f}) becomes identical to that of the XY chain.
In fact, if we introduce
 \begin{equation}
    \eta_{2j-1}^{(1)}  =  \sigma_{2j-1}^x \sigma_{2j}^x  \ , \
    \eta_{2j}^{(1)}  =  \sigma_{2j}^y \sigma_{2j+1}^y  \ , \
    \zeta_{2j-1}^{(1)}  =  \sigma_{2j}^x \sigma_{2j+1}^x  \ , \
    \zeta_{2j}^{(1)}  =  \sigma_{2j+1}^y \sigma_{2j+2}^y,
   \label{XY} 
 \end{equation}
then the operators $ \eta_j^{(1)}  $ and $ \zeta_k^{(1)} $ 
satisfy the same relations as $ \eta_j $ and $ \zeta_k $, 
and 
 $ K_0 \sum_j ( \eta_{2j-1}^{(1)}  + \zeta_{2j-1}^{(1)} ) $
and
 $ K_1 \sum_j ( \zeta_{2j}^{(1)} + \eta_{2j}^{(1)} ) $
 are nothing but the interactions of the XY chain. 
 Thus the model (\ref{Hamiltonian}) with $ K_2 = K_{-1} = 0 $ 
and the XY chain obey the same algebraic relation,
and therefore result in the same free energy.
\section{Gapless condition and the phase diagram} 

Let us consider the gapless condition that $ A(q) = |L(q)|^2 = 0 $ with some $q$.
The condition $ L(q) = 0 $ is equivalent to
 \begin{equation}
   \alpha_2 z^3 + \alpha_1 z^2 - z + \alpha_{-1} = 0
   , 
  \label{gap} 
 \end{equation}
where
 $ z = e^{iq} $, $ \alpha_{-1} = K_{-1}/K_0 $ , $ \alpha_1 = K_1/K_0 $ , $\alpha_2 = K_2/K_0 $.
Thus the model is gapless if and only if the algebraic equation (\ref{gap}) has a root 
that belongs to $ C_u  = \{ z \in \mathbb{C} \ | \ |z| = 1 \ \} $.
We find that the condition is equivalent to ($ X1 $) or ($ X2 $) or ($ X3 $), where
 \begin{align}
  & (X1) \ \ \alpha_1 = -\alpha_{-1} + ( \alpha_2 -1 ) \ \ {\rm or } \ \ \alpha_1 = -\alpha_{-1} - ( \alpha_2 -1 ) , 
   \notag \\
  & (X2) \ \ \alpha_{-1} \neq 0 \ \ {\rm and } \ \ \alpha_1 
                =
                \alpha_{-1} - \frac{ \alpha_2( 1+\alpha_2 ) }{ \alpha_{-1} } \ \ {\rm and } \ \ (1+\alpha_2)^2 - 4\alpha_{-1}^2 < 0,  
  \notag \\
  & (X3) \ \ \alpha_{-1} = 0 \ \ {\rm and } \ \ -2<\alpha_1 < 2 \ \ {\rm and } \ \ \alpha_2 = -1. 
 \end{align}
A derivation is given in Appendix A. These ($X1$) - ($X3$) form the boundary of 
each phase in the diagram shown in Figure 1 and 2.

 \begin{figure} 
  \centering
 \includegraphics[width=14cm]{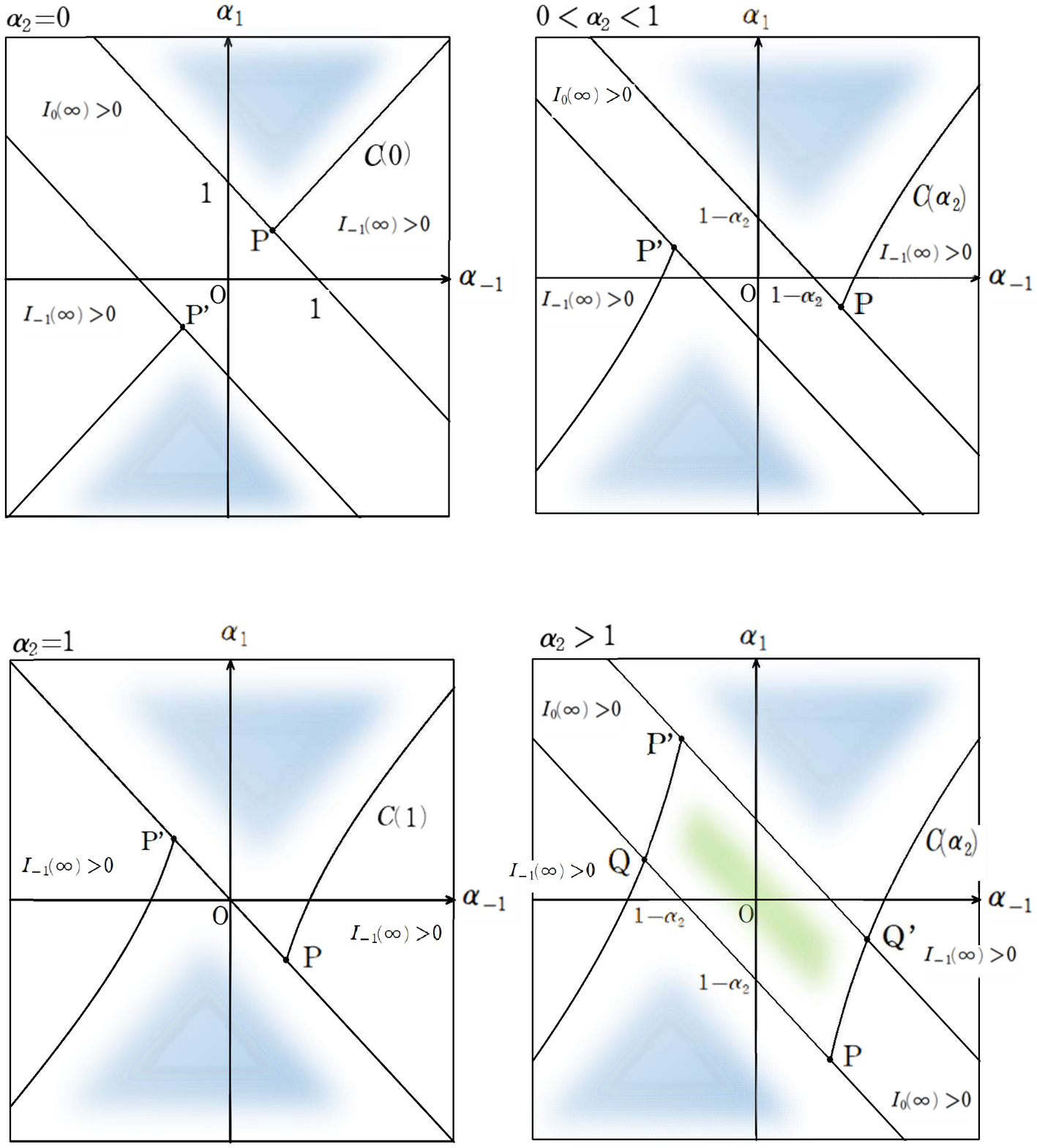}
  \caption{ 
     The phase diagram on $ \alpha_{-1}$ - $ \alpha_1 $ plane for each $ \alpha_2 \geq 0 $.
     Two lines denote
                   $ \alpha_1 = -\alpha_{-1}  \pm ( \alpha_2 -1 ) $, 
     and 
                   $ C( \alpha_2 ) $
    denotes the curve 
                   $ \displaystyle \alpha_1 = \alpha_{-1} - \frac{ \alpha_2(1+\alpha_2) }{ \alpha_{-1} } $.
                   P$ \displaystyle (\frac{ 1 + \alpha_{2} }{ 2 }, \frac{ 1-3\alpha_{2} }{ 2 }) $, 
                   Q$ \displaystyle ( -\alpha_{2}, 1) $, 
                   P'$ \displaystyle ( -\frac{ 1+\alpha_{2} }{ 2 }, -\frac{1-3\alpha_{2}}{2}) $, 
               and 
                  Q'$ \displaystyle (\alpha_{2}, -1) $ 
               are the multicritical points. Regions characterized by 
                  $ I_1(\infty) > 0 $ and $ I_2(\infty) > 0 $ 
               are colored by blue and green, respectively. 
               The central charge on each critical line is derived in section 5.
              }
 \end{figure}

 \begin{figure} 
  \centering
  \includegraphics[width=14cm]{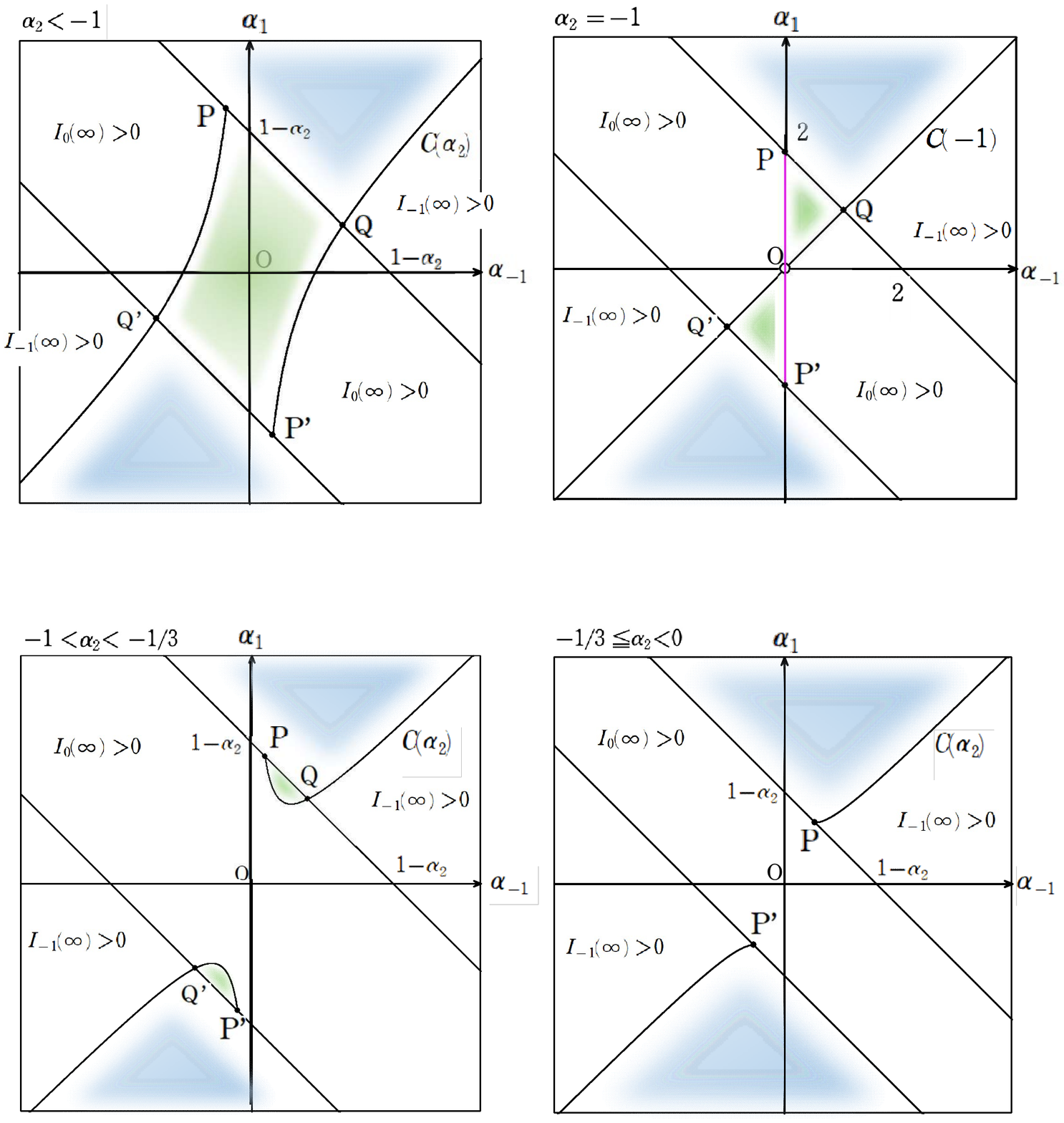}
  \caption{
         The phase diagram for each $ \alpha_2 < 0 $. 
         The red segment corresponds to $ (X3) $.
         In the case $ \alpha_{2} = -1 $, the point O$(0, 0)$  
         does not satisfy $(X2)$ but satisfy $(X3)$. 
             }
 \end{figure}

\section{Correlation function and the ground state phase transition} 

Next let us investigate a ground-state correlation function and its asymptotic behavior.
The correlation function we consider is
  \begin{equation}
     I_1(2n) 
        = \langle \ \sigma_{2j}^x \sigma_{ 2j+2n }^x \ \rangle_0
       ,
   \label{co1} 
  \end{equation}
where $\langle \ \ \rangle_0 $ is the ground-state expectation.
In the thermodynamic limit, we obtain the exact expression 
  \begin{equation}
    I_1(2n) =
       \left( \frac{2}{i} \right )^n
           \begin{vmatrix}
                 D(1) & D(2) & \cdots & D(n) \\
                 D(0) & D(1) & \cdots & D(n-1) \\
             \vdots & \vdots  & \ddots & \vdots \\
                D(2-n) & D(3-n) & \cdots & D(1)
           \end{vmatrix}
     ,  
   \label{to} 
  \end{equation}
where
 \begin{equation}
        D(r) = \frac{i}{4 \pi } \int_{- \pi}^{\pi} d q 
                     \left( e^{-iqr} \frac{ \sqrt{A(q)} }{ L(q)^{\ast} } \right)
                 .
  \label{co2} 
 \end{equation}
A derivation of (\ref{to}) and (\ref{co2}) is given in Appendix B.

The determinant (\ref{to}) is a Toeplitz determinant, i.e. $ ( j , k ) $ elements  depend
only on $ k-j=r $ . We can therefore apply the Szeg\"o's theorem to obtain the 
asymptotic limit $ n \to \infty $ of the determinant.
Let  $ C(r)=\frac{2}{i}D(r+1) $ , then (\ref{to}) is expressed as
  \begin{equation}
      \begin{vmatrix}
            C(0)  & C(1) & \cdots & C(n-1) \\
            C(-1) & C(0) & \cdots & C(n-2) \\
           \vdots  & \vdots      & \ddots & \vdots \\
            C(1-n) & C(2-n) & \cdots & C(0)
       \end{vmatrix}
     .
    \label{to1} 
  \end{equation}
Let us introduce $ f(p) $ by the relation
 \begin{equation}
    C(r) = \frac{1}{2 \pi} \int_{ -\pi }^{\pi} e^{-ipr} f(p) d p
     .
   \label{C} 
 \end{equation}
Then the Szeg\"o's theorem says that the asymptotic form of (\ref{to}) is
 \begin{equation}
     \lim_{ n \to \infty }
         \frac{ \text{det} ( \{ C(r) \} ) }{ \lambda^n } 
           = 
         \text{exp} \left( \sum_{n=1}^{\infty} n g_n g_{-n} \right)
   ,
  \label{seg} 
 \end{equation}
where
 \begin{equation}
     g_n  =  \frac{1}{ 2 \pi }  \int_{ -\pi }^{\pi} e^{-ipn}  \log f(p)  dp
  \label{defgn} 
 \end{equation}
and $ \lambda $ = exp $ g_0 $ . 
Note that $ g_{-n} = -g_n $.
We obtain, from (\ref{co2}) and (\ref{C}) that
 \begin{equation}
      f(p)  =  \frac{ \sqrt{ L(p) L(p)^{\ast} } }{ L(p)^{\ast} } e^{-ip}  
            =   \sqrt{  \frac{ e^{ -ip} L(p) }  { e^{ip} L(p)^{\ast} }  } ,
 \end{equation}
thus
 \begin{equation*}
     \log f(p) =
                   \frac{1}{2} \log (  \alpha_{-1} e^{-2ip}  -  e^{-ip}  +  \alpha_1  +  \alpha_2 e^{ip}  )
                - \frac{1}{2} \log (  \alpha_{-1} e^{2ip}    -  e^{ip}   +  \alpha_1  +  \alpha_2 e^{-ip}  ).
 \end{equation*}
With the use of the equality
 \begin{equation}       
       \alpha_{-1} e^{-2ip}  -  e^{-ip}   + \alpha_1  +  \alpha_2 e^{ip} 
         =
           A ( 1 + a_1 e^{ip} ) ( 1 + a_2 e^{-ip} ) ( 1 + a_3 e^{-ip} )
   , 
  \label{c1} 
 \end{equation}
 \begin{equation}
      A  =  (-t) \alpha_2 ,  \ \
      a_1 =  \frac{-1}{t} ,  \ \ 
      a_2 + a_3  =  \frac{ \alpha_1}{ \alpha_2 } + t ,  \ \
      a_2 a_3  =  \frac{-1}{t} \frac{ \alpha_{-1} }{ \alpha_2 }, 
  \label{c2}  
 \end{equation}
where $ t $ is a real solution of the equation 
 $ \alpha_{-1}  -  t  +  \alpha_1 t^2  +  \alpha_2 t^3  = 0 $,
and the factor $A$ is independent of $p$, we obtain 
 \begin{align}
      \log f(p) &= \frac{1}{2} \left( \log ( 1+ a_1 e^{ip} ) + \log ( 1+a_2 e^{-ip} ) + \log ( 1+ a_3 e^{-ip} ) \right) 
   \notag \\
                  & \ \   -\frac{1}{2} \left( \log ( 1+ a_1 e^{-ip} ) + \log ( 1+a_2 e^{ip} ) + \log ( 1+ a_3 e^{ip} ) \right).
  \label{logf} 
 \end{align}
Then we consider the following four cases:
  \begin{itemize}
      \item[ ($ \rm i $) ] $ |a_1|<1 $ and $ |a_2|<1 $ and $ | a_3|<1 $,
      \item[ ($ \rm ii $) ] $ |a_1|>1 $ and $ |a_2|>1 $ and $ | a_3|<1 $,
      \item[ ($ \rm iii $) ] $ |a_1|>1 $ and $ |a_2|<1 $ and $ | a_3|>1 $,
      \item[ ($ \rm iv $) ]  the other cases.   
  \end{itemize}
{ \bf case($ \rm i $) } Let $ |b|<1 $, then
  \begin{equation}
      \frac{1}{2\pi}  \int_{- \pi}^{\pi} e^{-inp} \log ( 1+be^{ip} ) dp
           = 
               \begin{cases}
                     \frac{(-1)^{n-1} }{n}  b^n  \ \  ( n > 0 )  \\
                     0  \ \ \ \  otherwise
               \end{cases}
    \label{b1} 
  \end{equation}
and
   \begin{equation}
      \frac{1}{2\pi}  \int_{- \pi}^{\pi} e^{-inp} \log ( 1+be^{-ip} ) dp
          =
            \begin{cases}
                     \frac{(-1)^{-n-1} }{-n}  b^{-n} \ \ ( n<0) \\
                     0 \ \ \ \ otherwise
            \end{cases}
      .
    \label{b2} 
  \end{equation}
From (\ref{defgn}), (\ref{logf}), (\ref{b1}) and (\ref{b2}), we obtain
  \begin{equation}
       g_0 = 0,  \ \ \   g_n = \frac{1}{2} \frac{(-1)^{n-1}}{n} ( a_1^n - a_2^n - a_3^n )  \ \  (n>0) 
.
  \end{equation}
Then from (\ref{seg}) we obtain
  \begin{equation}
      \lim_{ n \to \infty }
              I_1(2n)
               =
                 \left[ 
                        ( 1-a_1^2 ) ( 1-a_2^2 ) ( 1-a_3^2 ) 
                        ( 1-a_1 a_2 )^{-2}  ( 1 - a_1 a_3 )^{-2} ( 1 - a_2 a_3 )^2
                 \right]^{ \frac{1}{4} }
      .
    \label{as1} 
  \end{equation}
{ \bf case($ \rm ii $) } Let $|b|>1$, then
  \begin{equation}
      \frac{1}{2\pi}  \int_{- \pi}^{\pi} e^{-inp}  \log ( 1+be^{-ip} ) dp
         = 
           \frac{(-1)^n}{n}
              +
                \begin{cases}
                     \frac{(-1)^{n-1} }{n}  \left( \frac{1}{b} \right )^n  \ \  ( n>0 )  \\
                     0  \ \ \ \  otherwise
                \end{cases}
   \label{b3} 
  \end{equation}
and
   \begin{equation}
      \frac{1}{2\pi} \int_{- \pi}^{\pi} e^{-inp} \log ( 1+be^{ip} ) d p
         = 
           \frac{(-1)^n}{-n}
             +
            \begin{cases}
                     \frac{(-1)^{-n-1} }{-n} \left( \frac{1}{b} \right)^n\ \ ( n<0) \\
                     0 \ \ \ \ otherwise
            \end{cases}
    .
   \label{b4} 
  \end{equation}
From (\ref{defgn}), (\ref{logf}), (\ref{b3}) and (\ref{b4}), we obtain
  \begin{equation}
      \lim_{ n \to \infty }I_1(2n)
         =
             \left[ \left( 1-\frac{1}{a_1^2} \right)
                      \left( 1-\frac{1}{a_2^2} \right)
                      \left( 1- a_3^2 \right)
                      \left( 1-\frac{1}{ a_1 a_2 } \right)^{-2}
                      \left( 1-\frac{ a_3 }{ a_1 } \right)^2
                      \left( 1-\frac{ a_3 }{ a_2 } \right)^{-2}
              \right]^{\frac{1}{4}}
    .
   \label{as2} 
  \end{equation}
The limit  $ \displaystyle \lim_{ n \to \infty } I_1(2n) $  in the case ($ \rm iii $)
is obtained from (\ref{as2}) with
 $ a_2 $ and $ a_3 $
replaced by
 $ a_3 $ and $ a_2 $, 
respectively. 
In the case ($ \rm iv $), it is easy to derive that
 $ \displaystyle \sum_{n=1}^{\infty} n g_n g_{-n} = -\infty $
and
 $ \displaystyle \lim_{n \to \infty} I_1(2n) = 0 $.
From (\ref{as1}) and (\ref{as2}), we find that
 $ \displaystyle I_1( \infty ) = \lim_{n \to \infty } I(2n) >0 $
if and only if $a_1$ , $a_2$ and $a_3$ satisfy
 ($ \rm i $) or ($ \rm ii $) or ($ \rm iii $). 

Next let us consider the ground state phase diagram.
From (\ref{c1}), we obtain 
  \begin{equation}
     \alpha_2 (e^{ip})^3 + \alpha_1 ( e^{ip} )^2 - e^{ip} + \alpha_{-1} 
      =
      A a_1 ( e^{ip} + \frac{1}{a_1}  )(  e^{ip} + a_2  )(  e^{ip} + a_3 )
   .
  \end{equation}
Therefore, $ -1/a_1 $ , $ -a_2 $ and $ -a_3 $ are the roots of the cubic equation
 \begin{equation}
      \alpha_2 z^3 + \alpha_1 z^2 - z + \alpha_{-1} = 0
   .
  \label{equ3}  
 \end{equation}
Hence ($ \rm i $) - ($ \rm iii $) is satisfied if and only if
 \begin{equation}
    { \rm one \ root \ of \ (\ref{equ3}) \ belongs \ to } \ D_O  \ 
    { \rm and \ the \ other \ two \ roots \ belong \ to } \  D_I
  ,
  \label{con} 
 \end{equation}
where
    $ D_O  =  \{ z \in \mathbb{C} \ | \ |z|>1 \} $
 and
    $ D_I  =  \{ z \in \mathbb{C} \ | \ |z| < 1 \} $.
The equation (\ref{equ3}) is nothing but the equation (\ref{gap})
( the same equation appears after (\ref{c2}) ).
Therefore the boundary of the region $ I_1(\infty) > 0 $ satisfies the gapless condition, i.e. 
the curves corresponding to the gapless condition form the boundary of the ground state phase transition
in terms of the correlation function (\ref{co1}).
About the other regions bounded by the gapless curves, we consider correlation functions

 \begin{eqnarray}
      I_{l}(2n)
        =\Big( \frac{2}{i} \Big)^{n}
           \Big\langle
                   \prod_{k=1}^{n}   \varphi_{2}(j+k-l)\varphi_{1}(j+k)
           \Big\rangle_{0}
   \hspace{0.8cm}
      (l=-1, 0, 2),
 \end{eqnarray}
namely 
 \begin{eqnarray}
    I_{-1}(2n) &=& \Big\langle
                          \prod_{k=1}^{n}
                             \sigma^{x}_{2(j+k+1)-3}
                             \sigma^{y}_{2(j+k+1)-2}1_{2(j+k+1)-1}
                             \sigma^{y}_{2(j+k+1)}
                             \sigma^{x}_{2(j+k+1)+1}
                        \Big\rangle_{0}
  \nonumber \\
                &=& \Big\langle
                             \sigma^{x}_{2j+1}\sigma^{y}_{2j+2}\sigma^{x}_{2j+3}
                         \Bigg(
                                 1_{2j+4}\cdots\cdots 1_{2(j+n)}
                         \Bigg)
                            \sigma^{x}_{2(j+n)+1}\sigma^{y}_{2(j+n)+2}\sigma^{x}_{2(j+n)+3}
                      \Big\rangle_{0},
  \\
  I_{0}(2n) &=& \Big\langle
                       \prod_{k=1}^{n}
                            ( -\sigma^{x}_{2(j+k)-1} \sigma^{z}_{2(j+k)} \sigma^{x}_{2(j+k)+1} )
                    \Big\rangle_{0}
           \nonumber \\
              &=& (-1)^{n}
                    \Big\langle
                       \sigma^{x}_{2j+1}
                         \Bigg(
                              \sigma^{z}_{2j+2}1_{2j+3} \sigma^{z}_{2j+4}1_{2j+5}
                                   \cdots\cdots 
                              \sigma^{z}_{2(j+n)-2}1_{2(j+n)-1} \sigma^{z}_{2(j+n)}
                         \Bigg)
                      \sigma^{x}_{2(j+n)+1}
                   \Big\rangle_{0},
  \\
   I_{2}(2n) &=& \Big\langle
                         \prod_{k=1}^{n}
                              (
                                  -\sigma^{x}_{2(j+k-2)}
                                    \sigma^{x}_{2(j+k-2)+1}
                                    \sigma^{z}_{2(j+k-2)+2}
                                    \sigma^{x}_{2(j+k-2)+3}
                                    \sigma^{x}_{2(j+k-2)+4}
                              )
                     \Big\rangle_{0}
  \nonumber \\
              &=& \Big\langle
                              \sigma^{x}_{2j-2} \sigma^{x}_{2j-1} \sigma^{y}_{2j}
  \nonumber \\
             && \hspace{1.5cm}
                      \cdot\:
                     \Bigg(
                           1_{2j+1} \sigma^{z}_{2j+2} 1_{2j+3} \sigma^{z}_{2j+4}
                        \cdots \cdots 
                           1_{2(j+n)-5} \sigma^{z}_{2(j+n)-4} 1_{2(j+n)-3}
                     \Bigg)
  \nonumber \\
            && \hspace{7.5cm}
                     \cdot\: 
                        \sigma^{y}_{2(j+n)-2} \sigma^{x}_{2(j+n)-1} \sigma^{x}_{2(j+n)}
                  \Big\rangle_{0}.
\end{eqnarray}
Similarly we find that
 \begin{align}
    &(a) \ I_{-1}( \infty ) >0 \
             { \rm if \ and \ only \ if \ all \ the \ roots \ of \ (\ref{equ3}) \ belong \ to } \ D_O, \\
    &(b) \ I_{0}( \infty ) >0 \ 
            { \rm if \ and \ only \ if } \notag \\
           & \ \ \ \ \ \ { \rm \ one \ root \ of \ (\ref{equ3}) \ belongs \ to } \ D_I  \ 
            { \rm and \ the \ other \ two \ roots \ belong \ to } \  D_O, \\
    &(c) \ I_{2}( \infty ) >0 \
             { \rm if \ and \ only \ if \ all \ the \ roots \ of \ (\ref{equ3}) \ belong \ to } \ D_I.
 \end{align}
%
%
The results are shown in Figure 1 and 2.

Note that the ground state phase diagram is already obtained
in [24] for $ K_2 = 0 $.  
It is easy to convince that the phase diagram in Figure 1 with $ \alpha_2 = 0 $ 
is identical to the diagram in [24]. 

\section{Central charge} 

Let us consider the central charges of the theories  
on the critical lines ($X1$) - ($X3$). 
The points P, Q, P' and Q' in Figure 1 and 2 
are the multicritical points 
and will be considered at the end of this section. 
Except for these points, 
first we consider ($X1$) in particular the case  
 $ \alpha_1 = -\alpha_{-1}-(\alpha_2-1) $. 
Then
 $ L(q) = K_0 e^{-iq} ( e^{iq}-1 ) g(e^{iq}) $,
where
 $ g(z) = \alpha_2 z^2 + (1-\alpha_{-1})z-\alpha_{-1} $,
and
 $ g(e^{iq}) \neq 0 $. 
Then, we obtain
 \begin{equation}
   \Lambda(q) = 2 | L(q) | = 4| K_0 | | \sin \frac{q}{2} | | g(e^{iq}) |.
 \end{equation}

When
 $ \alpha_2 \neq 1 $,
we obtain the dispersion relation
 $ \Lambda(q) \simeq 4| K_0 | | p |C $, 
where $C=| g(1) |$ and 
 $\displaystyle p=\frac{2k-1}{N}\pi=\frac{1}{2}\frac{2k-1}{M}\pi=\frac{q}{2}$. 
Thus the conformal invariant normalizatin of the Hamiltonian is obtained[28] 
from the condition
 \begin{equation}
 4| K_0 | C = 1
  . 
  \label{cftnormalcon1} 
 \end{equation}
The term proportional to $1/N $ 
in the $1/N$ expansion of the ground-state energy
 $ E_0 = - \sum_{ 0<q<\pi } \Lambda(q) $
is obtained from 
 \begin{eqnarray}
  -4| K_0 | C \sum_{ 0<q<\pi } | \sin \frac{q}{2} |
    &=& 
  -4| K_0 | C \frac{1}{\displaystyle 2\sin \frac{ \pi }{N} }
\nonumber\\
&=& 
-2| K_0 | C \frac{N}{\pi} -2| K_0 | C\frac{1}{6} \frac{\pi}{N} + O( \frac{1}{N^{3}} ), 
 \end{eqnarray}
where we have used the relation 
 \begin{eqnarray}
   \sum_{k=1}^{M/2}\Big| \sin\frac{1}{2}\frac{2k-1}{M}\pi\Big|
     =
   \frac{1}{\displaystyle 2\sin \frac{ \pi }{2M} }.
 \end{eqnarray}
The second term  
  $ -2 | K_0 | C \pi / 6N $
is, from (\ref{cftnormalcon1}), equal to
 $- \pi /12N $. 
From the condition[29][30] that this term should be equal to
 $ -c \pi /6N$,
we obtain the central charge $c=1/2$. 
On the other critical line 
 $ \alpha_1 = -\alpha_{-1}+(\alpha_2-1) $, 
we similarly obtain $ c=1/2 $. 

When $ \alpha_2 =1 $, 
we obtain 
 $ L(q) = K_0 e^{-iq} ( e^{2iq}-1 )( e^{iq}- \alpha_{-1} ) $. 
We find two elementary excitations 
 $ v_{1}|q| $ and $ v_{2} | q - \pi | $. 
If $ \alpha_{-1} \neq 0 $, then $v_{1} \neq v_{2} $.  
If $\alpha_{-1} = 0 $, then $ v_{1} = v_{2} $, 
and considering again the term proportional to $1/N$, 
we obtain $c=1$.

Next, let us consider $(X2)$ and $(X3)$. 
The equation (\ref{equ3})
has a real root $B$ and two imaginary roots $ e^{iq_0} $ and $ e^{-iq_0} $,
where $ |B| \neq 1 $ and $ 0<q_0<\pi$.
Then we obtain 
 $ L(q) = K_0 \alpha_2 e^{-iq} (e^{iq}-B)(e^{iq}-e^{iq_0})(e^{iq}-e^{-iq_0}) $ 
and 
 \begin{equation}
   \Lambda(q) = 4|K_0 \alpha_2 | \sqrt{ 1+B^2 -2B \cos q} \ |\cos q_0 - \cos q |
   .
  \label{excitation} 
 \end{equation}
From (\ref{excitation}), we find that 
the conformal invariant normalization is
 \begin{equation}
   8 | K_0 \alpha_2 | \sqrt{ 1+B^2 -2B \cos q_0} \ \sin q_0 =1
   .
  \label{cftnormalcon2} 
 \end{equation}
The term proportional to $1/N$ is obtained from  
 \begin{equation}
    - 4|K_0 \alpha_2 | \sqrt{ 1+B^2 -2B \cos q_0} \
    \frac{ \sin q_0 }{\displaystyle \sin \frac{ \pi}{ M } }
    .
   \label{cftterm} 
 \end{equation}
From (\ref{cftnormalcon2}) and (\ref{cftterm}), 
we find that $ - \pi / 6N $ should be equal to $ -c \pi /6N$, 
and obtain the central charege $c=1$.

At the point Q (at the point Q'),  
two linear elementary excitations 
  $ v_{1} |q| $ and $ v_{2} |q-q_{0}| $ where $q_{0} \neq 0, 2\pi/3  $
 ( $ v_{1} |q-\pi| $ and $ v_{2} |q-q_{0}| $ where  $ q_{0} \neq \pi, \pi/3 $ ) 
appear, but
  $ v_{1} \neq v_{2} $. 

At the point P (at the point P'),  
then
 $ q_{0}=0 $ ($q_{0} = \pi $), 
and we find excitations proportional to 
 $ |q|^{2} $ (proportional to $ |q-\pi|^{2} $ ) 
when
 $ \alpha_{2} \neq 1, -1/3 $, 
and find excitations proportional to 
 $|q|^{2}$ and $|q-\pi|$ (proportional to $|q-\pi|^{2}$ and $|q|$) 
when 
 $ \alpha_{2} = 1 $, 
and an excitation proportional to 
 $ |q|^{3} $ (proportional to $ |q-\pi|^{3} $) when
 $ \alpha_{2} = -1/3 $.

\section{Symmetry and Generalizations} 

Let us first consider the symmetry of the phase diagram. 
When $ K_{0} \neq  0 $ and $ \alpha_{2} = 0 $, the gapless condition is simplified 
and expressed as 
 \begin{eqnarray} 
  &&
  (X1) 
  \hspace{0.5cm}
    (-\alpha_{1}) = -(-\alpha_{-1})\pm 1, 
  \nonumber\\
  &&
  (X2) 
  \hspace{0.5cm}
    (-\alpha_{-1}) \neq 0, 
  \hspace{0.3cm}
    (-\alpha_{1}) = (-\alpha_{-1}), 
  \hspace{0.3cm}
   1 - 4(-\alpha_{-1})^{2} < 0, 
  \nonumber\\
  &&
  (X3) 
  \hspace{0.5cm}
    {\rm cannot\:\: be\:\: satisfied}. 
  \label{cond-alpha}
 \end{eqnarray}
When $K_{1}\neq 0$ and $\alpha_{-1}=0$, 
let us consider the variables 
 $\beta_{-1}=K_{-1}/K_{1}$, 
 $\beta_{0}=K_{0}/K_{1}$, 
and 
 $\beta_{2}=K_{2}/K_{1}$, 
i.e. 
the normalization by $K_{1}$ instead of $K_{0}$. 
In this case the condition is expressed as 
 \begin{eqnarray}
  &&
  (X1) 
  \hspace{0.5cm}
    \beta_{2}=-(-\beta_{0})\pm 1, 
  \nonumber\\
  &&
  (X2) 
  \hspace{0.5cm}
     {\rm cannot\:\: be\:\: satisfied}. 
  \nonumber\\
  &&
  (X3) 
  \hspace{0.5cm}
    \beta_{-1}= 0, 
  \hspace{0.3cm}
    \beta_{2}=(-\beta_{0}), 
  \hspace{0.3cm}
    (-\beta_{0})<-1/2 \:\:{\rm or\:\:} 1/2<(-\beta_{0}).
  \label{cond-beta}
 \end{eqnarray}
From (\ref{majo}), we find the natural coupling constants of this model are 
  $K_{-1}$, $-K_{0}$, $K_{1}$, $K_{2}$. 
Thus when we consider the shift of indices of $K_{j}$ 
from $j$ to $j+1$, 
the correspondences of $\alpha_{j}$ and $\beta_{j}$ are  
\begin{eqnarray}
    (-\alpha_{-1})=\frac{K_{-1}}{-K_{0}}
   \mapsto
   \frac{-K_{0}}{K_{1}}=(-\beta_{0}),
 \hspace{0.5cm}
    (-\alpha_{1})=\frac{K_{1}}{-K_{0}}
   \mapsto
   \frac{K_{2}}{K_{1}}=\beta_{2}.
 \label{alpha-beta}
\end{eqnarray}
When we replace $ \alpha_j $ with corresponding $ \beta_j $ in (\ref{cond-alpha}), 
we find that the condition (\ref{cond-alpha}) becomes (\ref{cond-beta}). 
In both cases, 
the phase diagrams are also identical to 
that of the XY chain with an external field. 

Next  we will consider a generalization of the results obtained in Section 2-5.
We solve the model (\ref{Hamiltonian}) using only the algebraic relations of the operators (\ref{2}). 
Hence the model (\ref{mo}) generated from the operators which satisfy (\ref{2})
 can be simultaneously diagonalized, and its string order parameter (\ref{co1}) yields
 the same phase diagram as shown in  Figure 1 and 2.   
For example, from the operators (\ref{XY}), the model
 \begin{equation}
   K_0 \sum_{j=1}^N \sigma_j^x \sigma_{j+1}^x 
  + K_1 \sum_{j=1}^N \sigma_j^y \sigma_{j+1}^y 
  + K_2 \sum_{j=1}^N \sigma_j^x \sigma_{j+1}^z  \sigma_{j+2}^z \sigma_{j+3}^x
  + K_{-1} \sum_{j=1}^N \sigma_j^y \sigma_{j+1}^z  \sigma_{j+2}^z \sigma_{j+3}^y
  \label{ex1} 
 \end{equation}
is generated from (\ref{mo}), its string order parameter (\ref{co1}) in this case is
 \begin{equation}
    I_1(n) = \langle \prod_{k=1}^n ( \sigma_{2(j+k)-1}^y \sigma_{2(j+k)}^y ) \rangle_0 
         = \langle \prod_{k=1}^{2n} \sigma_{2j+k}^y \rangle_0
     ,
   \label{st1} 
 \end{equation}
and this model results in the free energy (\ref{f}) 
and the phase diagram shown in Figure 1 and 2. 

We introduce, in this paper, two series of operators 
$\{\eta_{j}\}$ and $\{\zeta_{j}\}$, 
which commute with each other. As a result, our model factorizes into two commutative
spin chains as discussed in [24] and [31].
One series of operators, however, 
can yield the free energy (\ref{f}) and the phase diagram shown in Figure 1 and 2. 
For example, the series of operators,
 \begin{eqnarray}
      \eta_{2j-1}^{(3)} = \sigma^{z}_{j}, 
   \hspace{0.6cm}
      \eta_{2j}^{(3)} = \sigma^{x}_{j} \sigma^{x}_{j+1}
  \label{zXYcluster}
 \end{eqnarray}
yields the Hamiltonian
 \begin{eqnarray}
    -\beta{\cal H}
   =
     K_{0}\sum_{j=1}^{N} \sigma^{z}_{j}
   + K_{1}\sum_{j=1}^{N} \sigma^{x}_{j} \sigma^{x}_{j+1}
   + K_{2}\sum_{j=1}^{N} \sigma^{x}_{j} \sigma^{z}_{j+1} \sigma^{x}_{j+2}
   + K_{-1}\sum_{j=1}^{N }\sigma^{y}_{j} \sigma^{y}_{j+1}
   .
   \label{ex2} 
 \end{eqnarray}
This Hamiltonian is composed of the XY interactions, 
an external field and the cluster interactions. 
The string order parameter (\ref{co1}) in this case is 
\begin{equation}
    I_1(n) = \langle \prod_{k=1}^n ( \sigma_{j+k-1}^x \sigma_{j+k}^x ) \rangle_0 
         = \langle \sigma_{j}^x \sigma_{j+n}^x \rangle_0
     ,
 \end{equation}
and this model also results in (\ref{f}) and the diagram in Figure 1 and 2. 
In case of both (\ref{XY}) and (\ref{zXYcluster}), 
the transformation (\ref{a}) results in the Jordan-Wigner transformation, 
and in case of (\ref{op1}) and (\ref{op2}), the transformation becomes 
(\ref{star1}) - (\ref{star4}).

Generally, 
an infinite number of operators that satisfy the condition (\ref{2}), 
including (\ref{XY}) and (\ref{zXYcluster}),  
are given in Table 1 and 2 in [24]. 
Our present results are universally valid  
for these infinite number of solvable spin chains. 
\section{Conclusion} 

In this paper, we obtain the exact solution of the model (\ref{Hamiltonian}), 
which is a cluster model that considers next-nearest-neighbor interactions
and two additional composite interactions. 
We introduce the series of operators (\ref{op1}) and (\ref{op2}), 
which satisfy the algebraic relation (\ref{2}). 
Then, we introduce the transformations (\ref{a}), 
and the model (\ref{Hamiltonian}) is diagonalized.
We derive the free energy (\ref{f}), 
consider a correlation function (\ref{co1}), 
and derive its exact expressions (\ref{to}) and (\ref{co2}).
We obtained the ground-state phase diagram, as shown in Figure 1 and 2, 
from the gapless condition, and from the asymptotic behavior of the correlation function, 
which is classified by the location 
of the roots of an algebraic equation (\ref{equ3}).  
We also derive the central charges of the corresponding CFT. 
The exact solution is obtained 
using only the algebraic relation (\ref{2}) of the interactions. 
Finally, we note that an infinite number of solvable models, 
generated from operators that satisfy (\ref{2}), 
yield exactly the same results obtained in this paper.
Our transformation can be regarded 
as an algebraic generalization of the Jordan-Wigner transformation. 
This paper provides a nontrivial example 
that cannot be solved by the Jordan-Wigner transformation 
but can be solved by our method.

\section*{Acknowledgements}   

The authors would like to thank Prof. T. Chikyu, Prof. K. Nomura, and Prof. H. Katsura for their comments. 
This work was supported by JSPS KAKENHI Grant Number JP19K03668.

\appendix 

\section{The gapless condition} 

Here we show the derivation of the condition that the equation (\ref{gap}) has a root which belongs to  $C_u$ if and only if ($X1$) or ($X2$) or ($X3$) is satisfied.
  
First, if the equation (\ref{gap}) has a root 1 or -1, then ($X1$) is satisfied.
Next, we assume that the equation (\ref{gap}) has imaginary roots which belong to $C_u$. 
When 
   $ \alpha_2 \neq 0 $ and $ \alpha_{-1} \neq 0 $,
 let $c$, $\bar{c} $ and $B$ be the roots of  the equation (\ref{gap}), where $c$, $\bar{c}$ 
are imaginary and $B$ is real.
Then from the cubic equation (\ref{gap}), we obtain
 \begin{equation}
     c + \bar{c} + B = -\frac{\alpha_1}{\alpha_2}  \ \ , \ \ c \bar{c} B = -\frac{\alpha_{-1}}{\alpha_2}.
 \end{equation}  
Thus $c$ and $\bar{c}$ are the imaginary roots of the quadratic equation
 \begin{equation}
     z^2  + ( \frac{\alpha_1}{\alpha_2} +B )z  -  \frac{\alpha_{-1}}{B \alpha_2} = 0.
  \label{qe} 
 \end{equation}
From (\ref{qe}), we obtain
 \begin{equation}
     ( \frac{ \alpha_1}{\alpha_2}+B )^2 +  \frac{4 \alpha_{-1} }{ B \alpha_2 } <0
  \label{imacon} 
 \end{equation}
and
 \begin{equation}   
     | c | = | \bar{c} | =  \sqrt{ -\frac{\alpha_{-1} }{ B \alpha_2} }.
  \label{ab} 
 \end{equation}
From $ |c | = |\bar{c}| = 1 $ and (\ref{ab}), we obtain 
 \begin{equation}
   B = -\frac{ \alpha_{-1} }{ \alpha_2 }.
  \label{B} 
 \end{equation}
By inserting (\ref{B}) to the equation (\ref{gap}), we obtain 
 \begin{equation}
  \alpha_1 = \alpha_{-1} - \frac{ \alpha_2 ( 1+ \alpha_2) }{ \alpha_{-1}}.
  \label{X21} 
 \end{equation}
On the other hand, From (\ref{imacon}), (\ref{B}) and (\ref{X21}), we find
 \begin{equation}
   ( 1+ \alpha_2 )^2 - 4 \alpha_{-1}^2 < 0.
  \label{X22} 
 \end{equation}
From (\ref{X21}) and (\ref{X22}) we obtain ($X2$).

When $ \alpha_2 \neq 0 $ and $ \alpha_{-1} = 0 $, the equation (\ref{gap}) becomes
   $ z ( \alpha_2 z^2 + \alpha_1 z  - 1 ) = 0 $.
Thus the imaginary roots of (\ref{gap}) are 
 \begin{equation}
      z = -\frac{ \alpha_1 \pm{i} \sqrt{ - 4 \alpha_2 - \alpha_1^2 } }{ 2 \alpha_2},
   \label{X3} 
 \end{equation}
where $ - 4 \alpha_2 - \alpha_1^2 > 0 $ is necessary. 
From  (\ref{X3}), we find 
 $ \displaystyle |z| = \frac{ \sqrt{ - \alpha_2 } } { | \alpha_2 | } $, 
and the condition ($X3$) is obtained.

When $ \alpha_2 = 0 $, the equation (\ref{gap}) becomes a quadratic equation
$ \alpha_1 z^2 - z + \alpha_{-1} = 0 $. 
From this equation, we obtain the condition that
 $ \alpha_1 = \alpha_{-1} $ and $ 1-4\alpha_{-1}^2 < 0 $, 
which is contained in ($X2$). 
Consequently, ($X1$) or ($X2$) or ($X3$) is a necessary condition.

In contradiction, when ($X1$) or ($X3$) is satisfied, it is easy to show that the equation
 (\ref{gap}) has a root which belongs to $C_u$.
When ($X2$) is satisfied, the equation (\ref{gap}) becomes
 $ ( \alpha_2 z + \alpha_{-1} ) \{ \alpha_{-1} z^2 -(\alpha_2 + 1 ) z + \alpha_{-1} \} = 0 $.
Thus (\ref{gap}) has imaginary roots
 \begin{equation}
   z = \frac{ (1+\alpha_2 ) \pm{i} \sqrt{ 4 \alpha_{-1}^2 - ( \alpha_2+1)^2} }{ 2 \alpha_{-1} },
 \end{equation} 
from which we find $ |z| = 1 $. Therefore, ($X1$) or ($X2$) or ($X3$) is sufficient.
\section{ Correlation functions} 

Here we derive the expression (\ref{to}) and (\ref{co2}) 
of the ground-state correlation function defined in (\ref{co1}).
The Hamiltonian is diagonalized by the canonical transformation
  \begin{gather}
     \xi_k(q) = \frac{1}{\sqrt{2A(q)}} ( -iL(q)C_k(q) + \sqrt{A(q)} C_{k+1}(q)) \ \ ( k=1, 3), \notag \\
     \xi_k(q) = \frac{1}{\sqrt{2A(q)}} ( -iL(q)C_{k-1}(q) - \sqrt{A(q)} C_{k}(q)) \ \ ( k=2, 4) 
   \label{g} 
  \end{gather}
as
  \begin{equation}
    -\beta \mathcal{H} =  \sum_{0<q<\pi} \Lambda(q) 
                        \{
                           \xi_1^{\dagger}(q) \xi_1(q) 
                           -\xi_2^{\dagger}(q) \xi_2(q)
                           +\xi_3^{\dagger}(q) \xi_3(q) 
                           -\xi_4^{\dagger}(q) \xi_4(q)
                         \}.
  \end{equation}
From (\ref{g}), We obtain
  \begin{equation}
      C_2^{\dagger}(q) C_1(q) = \frac{i}{2} \frac{ \sqrt{A(q)} }{ L(q) }
                    ( 
                      \xi_1^{\dagger}(q) \xi_1(q) 
                        +\xi_1^{\dagger}(q) \xi_2(q)
                        -\xi_2^{\dagger}(q) \xi_1(q) 
                        -\xi_2^{\dagger}(q) \xi_2(q) 
                     ).
  \label{c2c1} 
  \end{equation}
On the other hand,
  \begin{equation}
      \sigma_{2j}^x \sigma_{2j+2n}^x = \prod_{k=1}^n 
                                                 \sigma_{2(j+k)-2}^x 1_{2(j+k)-1} \sigma_{2(j+k)}^x
  \end{equation}
and from (\ref{et})
 $ \sigma_{2j}^x 1_{2j+1} \sigma_{2j+2}^x = \frac{2}{i} \varphi_2(j) \varphi_1(j+1) $,
therefore
  \begin{equation}
     I_1(2n) = \left( \frac{2}{i} \right)^n   
                \left\langle 
                     \prod_{k=1}^n \varphi_2(j+k-1) \varphi_1(j+k) 
                \right\rangle_0 .
  \end{equation}
Using the Wick's theorem, $I(2n) $ is expressed as
  \begin{equation}
     I_1(2n) = \left( \frac{2}{i} \right)^n
                \sum_P \text{sgn}(P) \prod_{k=1}^n   
                  \left\langle 
                      \varphi_2(j+k-1) \varphi_1(P(j+k)) 
                  \right\rangle_0
                  ,
   \label{r} 
  \end{equation}
where $P$ are the permutations of the indices. The sum (\ref{r}) is the determinant
  \begin{equation}
     \left( \frac{2}{i} \right)^n
        \begin{vmatrix}
            G_{j,j+1}    & G_{j,j+2}    & \cdots & G_{j,j+n} \\
            G_{j+1,j+1} & G_{j+1,j+2} & \cdots & G_{j+1,j+n} \\
            \vdots     & \vdots      & \ddots & \vdots \\
            G_{j+n-1,j+1} & G_{j+n-1,j+2} & \cdots & G_{j+n-1,j+n}
        \end{vmatrix}
        ,
     \label{det3} 
  \end{equation}
where $ G_{l,m} = \langle \varphi_2(l) \varphi_1(m) \rangle_0 $.
Because of  the translational invariance, (\ref{det3}) is expressed as (\ref{to}).
From (\ref{c2c1}) and the fact that the ground-state correlation functions satisfy
$ \langle \xi_1^{\dagger} (q) \xi_1 (q) \rangle_0 = 1$,
$ \langle \xi_2^{\dagger} (q) \xi_2 (q) \rangle_0 
  =
   \langle \xi_2^{\dagger} (q) \xi_1 (q) \rangle_0
  =
   \langle \xi_1^{\dagger} (q) \xi_2 (q) \rangle_0
  = 0
$, 
we obtain
  \begin{align}
    G_{j,j+r}
         &= \frac{1}{M} 
                \sum_{0<q<\pi} 
                   [ e^{iqr} \langle C_2^{\dagger}(q) C_1(q) \rangle_0
                       + e^{-iqr} \langle C_2(q) C_1^{\dagger}(1) \rangle_0 ] \notag \\
         &= \frac{1}{M} 
              \frac{i}{2}
                 \sum_{0<q<\pi}
                   \left( 
                      e^{iqr} \frac{ \sqrt{A(q)} }{ L(q) } + e^{-iqr} \frac{ \sqrt{A(q)} }{ L(q)^{\ast} }
                    \right)
         .
    \label{D} 
  \end{align}
In the thermodynamic limit, (\ref{D}) results in (\ref{co2}).

\end{document}